\documentclass[fleqn,10pt]{wlscirep}
\usepackage{pgfplots}
\usetikzlibrary{calc,arrows.meta}
\def\flushboth{%
	\let\\\@normalcr
	\@rightskip\z@skip \rightskip\@rightskip
	\leftskip\z@skip
	\parindent 1.5em\relax}
\title{Towards Picogram Detection of Superparamagnetic Iron-Oxide Particles Using a Gradiometric Receive Coil}
\author[1,*]{Matthias Graeser}
\author[2,3]{Tobias Knopp}
\author[2,3]{Patryk Szwargulski}
\author[1]{Thomas Friedrich}
\author[1]{Anselm von Gladiss}
\author[4]{Michael Kaul}
\author[5]{Kannan M Krishnan}
\author[4]{Harald Ittrich}
\author[4]{Gerhard Adam}
\author[1]{Thorsten M. Buzug}
\affil[1]{Institute of Medical Engineering, University of L\"ubeck, L\"ubeck, Germany}
\affil[2]{Section for Biomedical Imaging, University Medical Center Hamburg-Eppendorf, Hamburg, Germany}
\affil[3]{Institute for Biomedical Imaging, Hamburg University of Technology, Hamburg, Germany}
\affil[4]{Department for Diagnostic and Interventional Radiology and Nuclear Medicine, Hamburg University of Technology, Hamburg, Germany}
\affil[5]{Department of Materials Science and Department of Physics, University of Washington, Seattle, USA}
\affil[*]{ma.graeser@uke.de}

\newcommand{\eff}{\rho}
\renewcommand{\vec}[1]{\mathit{\mathbf{#1}}}

\usepackage{amsmath}
\usepackage{textcomp}
\usepackage{ngerman}%
\usepackage{pgfplots}

\begin{abstract}
Superparamagnetic iron-oxide nanoparticles can be used in a variety of medical applications like vascular or targeted imaging. Magnetic particle imaging (MPI) is a promising tomographic imaging technique that allows visualizing the 3D nanoparticle distribution concentration in a non-invasive manner. The two main strengths of MPI are high temporal resolution and high sensitivity. While the first has been proven in the assessment of dynamic processes like cardiac imaging, it is unknown how far the detection limit of MPI can be lowered. Within this work, we will present a highly sensitive gradiometric receive-coil unit combined with a noise-matching network tailored for the measurement of mice. The setup is capable of detecting 5~ng of iron \textit{in vitro} at 2.14~sec acquisition time. In terms of iron concentration we are able to detect 156~\textmu g/L marking the lowest value that has been reported for an MPI scanner so far. \textit{In vivo} MPI mouse images of a 512~ng bolus at 21.5~ms acquisition time allow for capturing the flow of an intravenously injected tracer through the heart of a mouse. Since it has been rather difficult to compare detection limits across MPI publications we propose guidelines improving the comparability of future MPI studies.
\end{abstract}

\begin{document}

\flushbottom
\maketitle
%
%
\thispagestyle{empty}

\section{Introduction}

Magnetic particle imaging (MPI) is a tracer based tomographic imaging methods capable of determining the 3D distribution of the tracer concentration \cite{Gleich2005,Knopp2012}. The tracer used in MPI is based on superparamagnetic iron-oxide (SPIO) nanoparticles consisting of an iron-oxide core that is surrounded by a shell preventing agglomeration \cite{Ferguson2015,Rogge2013,Weizenecker2012}. SPIONs have a long history of being flexible imaging biomarkers that can be used in biomedical applications. SPIONs can be used in magnetic resonance imaging (MRI), where they mostly provide a negative contrast, making it difficult distinguishing them from other negative contrast, e.g. flow, air, iron containing metabolites or even heterogeneous tissues. MPI has to the contrary a positive contrast and in turn allows for quantitative and background-free imaging of SPIONs.

MPI has a variety of potential medical applications \cite{Haegele2012c}. In particular, it is capable of imaging dynamic processes as required for cardiovascular applications \cite{Haegele2014}. The first \textit{in vivo} MPI experiment published in \cite{Weizenecker2009} revealed structures of a beating mouse heart during the in-flow of an intravenously injected SPIO bolus. Using long circulating tracers it is even possible to image the vessel tree over several hours \cite{khandhar2015tuning} making MPI very suitable for interventional image guided procedures \cite{Haegele2013,salamon2016magnetic,haegele2016multi} where nowadays the digital subtraction angiography (DSA) is clinically used. The DSA applies ionizing radiation and a potentially nephrotoxic iodine tracer that needs to be regularly reinjected due to dilution in the blood circulation and limited  half-life times of few minutes. 

Cell labeling and cell tracking are further very important applications for MPI \cite{Lindemann2014} that potentially  allow for direct imaging of cell movements in vivo. It has been shown that MPI can track the long-term fate of \textit{in vivo} neural cell implants with high image contrast  \cite{zheng2015magnetic}. It was also shown that cells implanted in different hemispheres of the brain can be resolved and that the precise number of cells can be quantified  \cite{bulte2015quantitative}. Furthermore, it has been suggested to substitute SPIONS for radioactive tracers in the sentinel lymph node biopsy, to carry out real-time image guided biopsy with a single-sided MPI device instead of scintegraphic imaging \cite{Ruhland2009}. 

In all of the proposed medical applications it is most crucial that MPI is sensitive enough to detect even very small amounts of SPIONs. Since there is a direct relation between the signal-to-noise-ratio (SNR) of the measurement signal and the spatial resolution \cite{knopp2011prediction}, a sensitive MPI scanner is the key to achieve high resolution MPI images.

The sensitivity of experimental MPI systems has been improved over the last years \cite{zheng2015magnetic, rahmer2013nanoparticle, Schmale2010, Schmale2010a, them2016sensitivity,them2016increasing}. The reported detection limits reach from 200 cells \cite{zheng2015magnetic} over 50~\textmu mol/L concentration within one voxel volume of 0.216~\textmu L \cite{rahmer2013nanoparticle} to 1~\textmu g iron content \cite{them2016sensitivity}. Unfortunately, these different detection limits are nearly impossible to compare as the systems differ in encoding schemes (field-free-point (FFP) and field-free-line (FFL), gradient strength (2.5\;T/m to 7\;T/m), reconstruction algorithm ($x$ space and frequency space reconstruction), drive-field strength (12\;mT to 20\;mT) and scan times (21\;ms to 20\;min). In addition, the methods to determine the detection limit differ considerably making these numbers even less comparable. 
Without reducing the relevance of the maximal sensitivity values for each imager, all findings must be related to scanner parameters and the experimental protocol. For the extrapolation to human use, scan times, drive field strength and the used frequencies have a direct impact on the specific absorption rate and the peripheral nerve stimulation limits \cite{Saritas2013a,Schmale2013,Schmale2015,Saritas2015}. Addressing these relations will make the results more comparable and easier to relate.

The purpose of this work is threefold: First, a highly sensitive MPI receive coil is developed that can be installed in a commercially available MPI scanner and improve its sensitivity.
The second aim of this work is to discuss all factors that influence the sensitivity in MPI. Based on that, we develop a procedure that makes it easy to compare different MPI scanners with respect to their sensitivity. The procedure is independent of the applied encoding scheme (FFL/FFP) and is also adapted to the bore diameter of the scanner.
Third, we will investigate the relationship between the detection limit in terms of total iron amount and the detection limit in terms of concentration. 

\section{Methods}
\subsection{Hardware Setup}

\subsubsection{Preclinical MPI Scanner}

\begin{figure}[tb]
\begin{center}
\includegraphics[width=\textwidth]{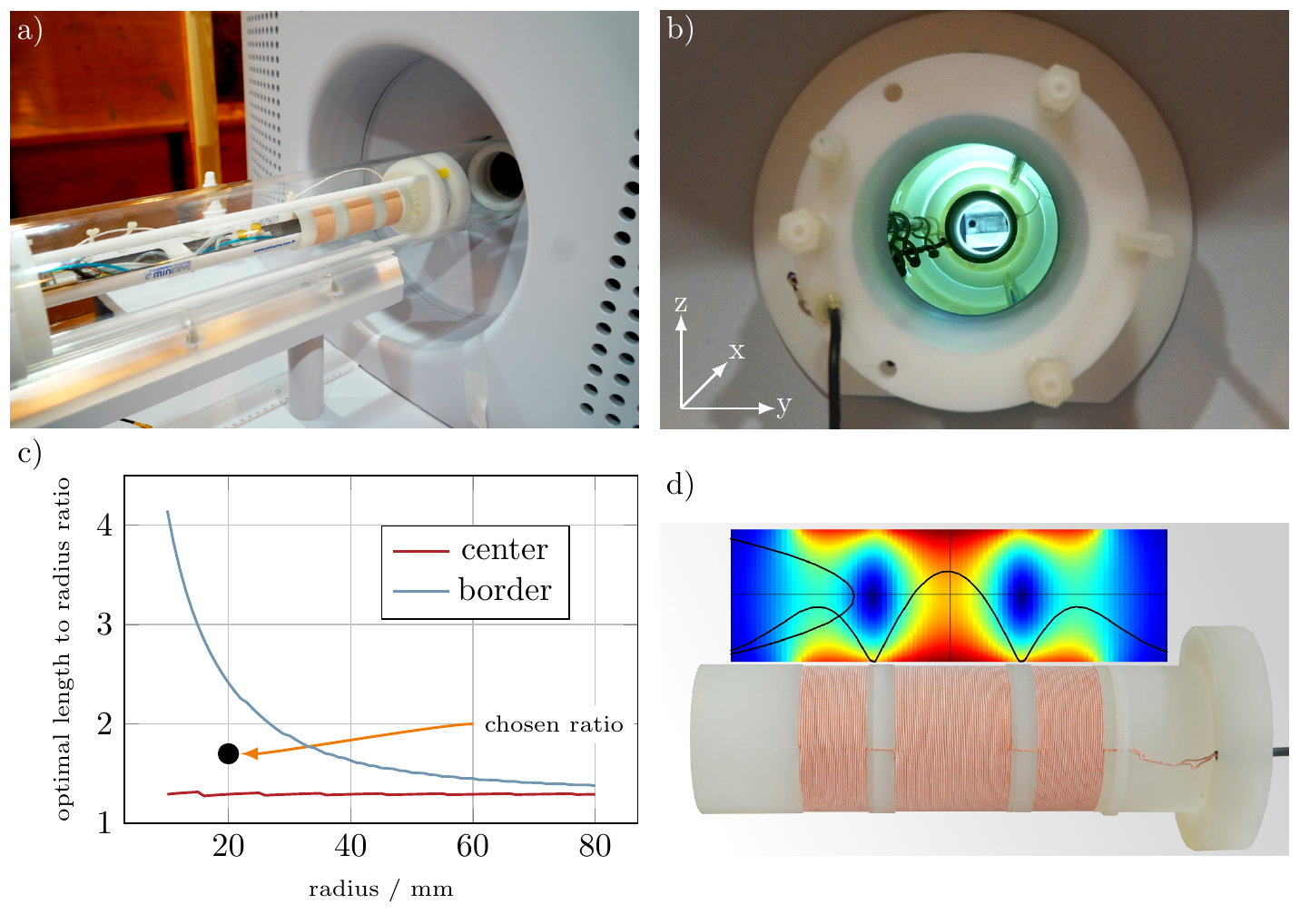}
\end{center}
\caption{a) Mouse insert placed in-front of the scanner bore. The receive coil is mounted within the insert, such that the mouse bed fits into. b) gradiometer coil mounted inside the scanner c) optimal coil length $l$ per radius $R$ at different radii for the FOV center and FOV border. For the aimed bore diameter of 40\,mm a compromise of 50 turns or 34\,mm ($l/R=1.7$)  was chosen (black dot). d) Gradiometric receive coil with a simulated field profile using numerical evaluation of the Biot-Savart law. }
\label{fig:MPIScanner}
\end{figure}

The preclinical MPI scanner (Bruker Biospin MRI GmbH, Karlsruhe, Germany )\cite{BrukerScanner} used in this work is shown in Fig. \ref{fig:MPIScanner} a). It has a 11.8~cm scanner bore diameter allowing to image small animals such as mice and rats. The scanner has an electromagnetic selection field generator adjustable between 0~T/m and 2.5~T/m. The selection field has an anisotropic gradient strength being twice as high in the $z$ direction (vertical) than in $x$ (colinear to scanner bore) and $y$ directions.
The FFP is steered along a 3D Lissajous curve using three orthogonal send coils that are wound around the scanner bore. They have a maximum amplitude of 14 mT and slightly different excitation frequencies with commensurable frequency ratios resulting in a repetition time of 21.5~ms. Within this work, the gradient strength of the selection field were set to 2.0 T/m and the drive field to 12\,mT leading to a field-of-view (FOV) size of $24\times 24\times 12$~mm$^3$. 

The MPI scanner is equipped with an animal support unit (Minerve, Esternay, France) providing anesthesia, temperature regulations, and monitoring of the respiratory rate. The unit is used for the examination of mice and can be mounted on both the MPI scanner and the MRI scanner (Bruker Clinscan, Karlsruhe, Germany) that is used for  anatomical reference imaging. The animal support unit is of a cylindrical shape and has an outer diameter of 38~mm.

\subsubsection{Combined Send/Receive Path}
The send coils of the scanner are not only used for excitation of the SPIO magnetization but  also  for signal detection. This is possible due to the narrowband excitation frequency range (24~kHz -- 27~kHz) that only slightly overlaps with the frequency range of the receive signal (0~Hz -- 1.25~MHz). 
The advantage of this type of receive chain is the very low noise of the transmit coil due to bigger copper cross sections. Additionally, the bore size is not reduced by a receive coil. This can lead either to a bigger bore size or to reduced power consumption due to a smaller transmit coil diameter. 
The disadvantage of a combined send/receive chain is the undamped direct feedthrough of higher harmonics from the send in the receive chain. A cancellation approach as introduced in \cite{Graeser2013} requires dedicated receive coils and therefore, its not possible for send/receive coils. This leads to higher demands on the purity of the transmit signal.

\subsubsection{Dedicated Receive Coil Path}
With its 11.8~cm bore diameter the preclinical MPI scanner is not fully optimized for the measurement of small animals. In this work, a dedicated receive coil is developed that is optimized for mice experiments.

While 3D drive-field MPI scanners usually use three independent receive coils for signal reception, a much simpler design consisting of a single receive coil is chosen. It can be motivated by theoretical findings outlined by Rahmer et al. \cite{Rahmer2012}, that a single receive channel is sufficient for the reconstruction of 3D Lissajous type MPI data. It was shown that the three orthogonal channels of a 3D Lissajous type scanner contain highly redundant data. Therefore,  discarding two of the three receive channels is expected to only marginally affect the image quality. In \cite{Szwargulski2017a} it has been experimentally shown that a removal of a receive channel leads to a loss in spatial resolution between 12\% and 22\%.

Since the receive coil of the scanner is fully coupled with the send chain, it has been necessary to develop an independent receive chain including receive coil, filter, and low noise amplifier (LNA). 
The developed insert consists of a receive coil and a mechanical construction to adapt the position of the coil as well as a tilt on the $y$ and the $z$ axes.
Since we observed a strong background signal in the installed coil, a gradiometric receive coil design of second order was chosen that is capable of reducing disturbing signals \cite{Karp1980,Tumanski2007}. In addition, it is capable of reducing the direct feedthrough of the send coils by approximately 60 dB. For non-gradiometric coils the stable background can be removed by subtracting an empty measurement. However, the gradiometric coil is even capable of reducing dynamic background signals that change during an experiment. We note that gradiometric receivers have been used in various MPI scanners and spectrometers before \cite{Graeser2013,tay2016high}.

The receive coil is connected to a fourth order butterworth filter canceling out the remaining fundamental frequencies of all three transmit channels. To digitize the signal it is amplified by a custom designed, battery powered LNA. The signal is then transferred into a differential signal via a 1:1 transformer. This differential signal is finally connected to the second digitizer channel of the MPI scanner. The results can then be compared between a first channel only reconstruction (preinstalled receive chain) and a second channel only reconstruction (gradiometric receive chain).

\message{***********\the\columnwidth}
\subsubsection{Noise Matching} \label{Sec:NoiseMatching}
The custom built LNA has an input noise level of $440\,\text{pV}/\sqrt{\text{Hz}}$ with 154\,pF input capacitance. The coil resistance is $1.42\,\Omega$. This leads to a coil noise level of $u_\text{Rx,rms}=154\,\text{pV}/\sqrt{\text{Hz}}$. To match these levels a noise matching transformer is introduced after the filter stage. The turn ratio of $\eta=\frac{N_s}{N_p}=2$ amplifies the signal and the coil noise by $\eta$. As the LNA is still noise dominant, this leads to an SNR increase by $\eta$ of the amplified signal. However, the transformer also introduces an impedance conversion ratio of $\eta^2$ leading to a bandwith reduction of $\eta$ due to \begin{align}
	f_\text{r}=\frac{1}{2 \pi \sqrt{LC}}.
\end{align} 
In this work, the design goal was a receive bandwidth of 300\,kHz to capture at least 10 harmonics of the excitation frequency and therefor guarantee a good image resolution. Consequently, it is not possible to use a higher turn ratio although coil noise dominance is not yet reached.
Under the assumption of constant field sensitivity for each turn of a coil, this noise matching is equivalent to an increase of coil turns. However, as each additional turn will be placed further away from the FOV center, the sensitivity will not scale linearly with the number of turns. Consequently noise matching is superior when the sensitivity increase by additional turns is smaller than the square-root of the inductance increase. The noise matching transformer is realized using a toroidal ferrite core of high permeability (Fair-Rite material 78, $35.5 \textrm{mm}\times23\text{mm}\times 12$ mm).  

\subsection{Sensitivity laws for optimizing receive coils}\label{Sec:SensitiviyLaws}
In contrast to the preinstalled receive coil, the developed gradiometer has a smaller accessible bore diameter. This directly affects the sensitivity of the system. To investigate this effect some general scaling laws are derived from approximative formulas for inductance and magnetic field of a short solenoid coil. The noise matching described in section \ref{Sec:NoiseMatching} allows for introducing an effective coil sensitivity $\eff$ under the assumption of an LNA noise dominated receive chain
\begin{align}
\eff=\frac{p_x(\vec{r})}{\sqrt{L}},
\label{Eq:CoilEff}
\end{align} with $\vec{p}=\left( p_x, p_y, p_z \right)^T$ being the coil sensitivity at unit current \cite{Knopp2012}, $\vec{r}$ being the spatial position and $L$ being the coil inductance.
For the center of the FOV the coil sensitivity $p_x$ can be calculated analytically by
\begin{align}
p_x(0)=\frac{N\mu_0}{l\sqrt{\frac{4R^2}{l^2}+1}}
\end{align} 
where $l$ is the coil length, $R$ is the coil radius and $N$ is the number of turns.
The inductance of a short air core solenoid can be calculated to be\cite{Wheeler1928}
\begin{align}
L=\frac{\mu_0 N^2 \pi R^2}{l+0.9 R}.
\end{align}
Then, the optimum length of the coil for the center position can be derived to be $l\approx1.3\cdot R$. Given this ratio, the radial dependency of the effective sensitivity $\eff$ is shown to be
\begin{align} \label{Eq:ScalingLaw}
\rho_\text{optim} & \propto \frac{1}{\sqrt{R^3}}
\end{align}
as is derived in the appendix (See section\ref{Appendix}). We note that the optimal ratio $\frac{l}{R}\approx1.3$ was only derived for the center. When considering a FOV around the center, the optimal ratio will change as it is discussed in the next section.

\subsubsection{Design of the Receive Coil}
The design of the  gradiometric receive coil is optimized in three stpdf. 
First, the inner diameter of the receive coil is set to the minimal value that it still fits over the animal support unit. Since the latter has a diameter of 38~mm the inner coil diameter is set to 40~mm. This gives about 1~mm on each side to ensure a flawless insertion of the animal bed into the receive coil insert.
For a fair comparison to the preinstalled coil one has to keep in mind that the receive sensitivity scales with the coil radius (see section \ref{Sec:SensitiviyLaws}). Thus, the bore reduction leads to a theoretical sensitivity increase of about 5.2. 

Second, the coil sensitivity $\mathbf{p}(\mathbf{r})$ within the FOV of $24\times 24\times 12$~mm$^3$ is simulated via a Biot-Savart approach. Additionally, the inductance of the coil is calculated. A litz wire of 0.68\,mm diameter (500 strands of $20$~\textmu m diameter) was chosen and the number of turns were  increased stepwise, going along with an increase in coil length. A single-layer coil was chosen since multiple layers are not as efficient as the alternative noise matching described in section \ref{Sec:NoiseMatching}.

The inductance of the receive coil limits the receive chain bandwidth due to a resonance with the capacitive elements within the receive chain. Usually, the total capacitance $C$ of the receiver network is dominated by the input capacitance of the LNA, whereas parasitic capacitances from the receive coil can be neglected. As bandwidth can be transferred into SNR via noise matching techniques (see section \ref{Sec:NoiseMatching}), the coil efficiency $\rho$ is used for optimization.
The maximum of $\rho$ marks the point where noise matching techniques become more effective than additional turns of the receive coil. For $R=20\,\text{mm}$, the optimal sensitivity for the center point is reached for 40 turns. However, with 40 turns the coil length is only 27.2\,mm leaving the corner of the $24$~mm FOV in the border region of the coil. The optimum for the border of the FOV on the $x$ axis (weakest sensitivity) is reached at 70 turns. As the efficiency of the receive coil drops only slightly besides the optimal values, a compromise of 50 turns is chosen for the manufactured coil. Figure \ref{fig:MPIScanner} c) shows the ratio of coil length and coil radius for the maximum value of $\rho$.

Third, the cancellation turns are determined. Here, the field profile of the drive field generator in bore direction is measured using a circular calibration coil. The coil was translated along the bore axis and the induced voltage in the coil was recorded using an oscilloscope. 
The total induced voltage in the receiver winding has been calculated via law of induction based on the field profile. The cancellation winding positions are then chosen in a way that the total induction in the cancellation turns matches the value of the receiver winding. A minimal distance of 6.8~mm is set between the receiving and cancellation turns to avoid a strong sensitivity reduction due to the cancellation turns.
The resulting coil consists of 50 receive turns and $2\times 29$ cancellation turns with the mentioned distance of 6.8\,mm in-between (see Fig. \ref{fig:MPIScanner} d)).

\subsection{Detection Limit Determination}

In this work, a simple but accurate method for the determination of the detection limit of an MPI scanning device is proposed. The most important guidelines of the proposal are:
\begin{enumerate}
\item \textbf{Use a dilution series}: In our experience, it is not suitable to use a highly concentrated sample and extrapolate the detection limit based on the SNR of the sample. The actual proof is only given, when a sample slightly over the detection limit can still be visualized. We recommend a dilution series with a exponential step factor of $\leq 2$. 
\item \textbf{Make the sample very small}: To make the procedure independent of the encoding scheme (FFL/FFP) and the applied gradient strength, we propose to make the sample as small as feasible. In these guidelines we recommend a 1~\textmu l sample. Determining the effectiveness of the encoding scheme can be an additional test after detection limit determination using a small point sample.
\item \textbf{Use reconstructed data}: It is fine to first analyze the raw data to determine a rough estimate of the detection limit, but it is crucial that this raw data is also be transformed into an actual image.
\item \textbf{Move the sample}: Due to the regularization of the reconstruction it may happen that the reconstructed images contain artifacts that are falsely identified as particles. In order to circumvent a misinterpretation, we propose to move the sample stepwise through the FOV to defined positions and verify if the movement can be identified in the reconstructed images.
\item \textbf{Relate the sensitivity to the measurement protocol and the scanner dimensions}. While it is possible to decouple the sensitivity estimation from the encoding scheme using a small sample, there are still various acquisition and scanner parameters that cannot be easily normalized. First, it is important to report the excitation field strength and frequency since both influence the SNR of the measured signal. Second, one has to report the used tracer material, since different SPIONs generate different MPI signals. Third, it is crucial to report the acquisition time. With increasing acquisition time the SNR of the measured signal improves in a root fashion. Finally, it is important to relate the sensitivity to the bore diameter of the scanner. The scaling law \eqref{Eq:ScalingLaw} then allows comparing different sensitivities.
\end{enumerate}

These guidelines make most of the experimental process comparable. However, due to different reconstruction techniques the influence of  frequency selection, regularization or deconvolution on the sensitivity limit can not as easily be excluded by general guidelines. 
\subsection{Experiments}

The following experiments are aligned with the proposed rules for determining the detection limit and give some additional insights.
\subsubsection{MPS Measurements}
The sensitivity scaling law \eqref{Eq:ScalingLaw} shows that the highest sensitivity can be achieved by very small receive coils. Magnetic particle spectroscopy (MPS) \cite{Biederer2009} may be used to forecast the sensitivity of trajectories and encoding schemes that are not realized today. The MPS used here has sandclock shaped receive coils with a maximum diameter of 1\,cm \cite{Graeser2017a}. To investigate the maximum sensitivity of the system, a 1:5 noise matching transformer is introduced in the receive chain. Then, a dilution series with LS-008 particles\cite{Ferguson2015,Kemp2016,Khandhar2017} with an absolute iron content ranging from 803~ng to 24.5~pg has been measured with a single drive field at 20~mT amplitude with a measurement time of 136\,ms. 

For Langevin particles, the signal strength is known to be proportional to the drive field amplitude \cite{Goodwill2010}. For anisotropic particles, a minimum drive field strength has to be applied in order to overcome the energy barrier \cite{Weizenecker2012,Graeser2015c, Graeser2015}. To investigate this fact an AC sweep was performed with the MPS varying the drive field amplitude from 2~mT to 20~mT. As particle sample, 5\,\textmu L of LS-008 was used. The measurement time has been 136~ms as well. 

\subsubsection{System Matrix Acquisition}
In order to reconstruct 3D Lissajous type MPI data, the method of choice is to determine the system matrix with a small calibration sample representing an image voxel. We use a sample of size $1\times 1\times 1$~mm$^3$ that is filled with undiluted LS-008 SPIONs (concentration of 5.14 g/L) \cite{Ferguson2015}. The sample is moved to $32\times 25 \times 13$ positions in 1~mm stpdf while continuously measuring the system response at each position. The measurement at a specific position is later assigned to the corresponding column in the system matrix that describes the mapping between measurement signal and particle distribution.

For the \textit{in vivo} measurements a system matrix with finer voxel spacing has been recorded to address for the fine structures in the cardiovascular system of a mouse. Therefore a 0.7~mm glass capillary was filled 0.7~mm high with undiluted LS-008 and was placed on the sample holder. The resulting SM contains $46\times36\times19$ voxel but suffers from less SNR due to the smaller amount of particles in the calibration sample.
  
All acquired system matrices as well as all later measurements have been corrected for the background of the system by taking dedicated empty measurements and subtracting them.

\subsubsection{Image Reconstruction}
For image reconstruction a former published framework is used \cite{Knopp2010e}. First, a frequency selection is performed where only those frequency components above a certain SNR threshold are taken into account. The SNR threshold parameter is varied across the dilution series. It is chosen based on the SNR of the system matrix and scaled with the dilution of the sampled object. The linear imaging equation is then solved using a regularized form of the iterative Kaczmarz algorithm \cite{Kaczmarz1937} that is known to converge rapidly in MPI \cite{Knopp2010e}. We use a single Kaczmarz iteration throughout all experiments. The regularization parameter is optimized based on visual inspection to provide the best image quality.

\subsubsection{Dilution Series}
To determine the detection limit of the developed receive coil a dilution series is measured with both the preinstalled receive path and the dedicated receive path at the same time. All samples feature the same volume of 1\,\textmu L that was placed in an microliter tube. 
Each step in the dilution series reduced the iron content by a factor of two, leading to iron contents ranging from 5.14\,\textmu g (undiluted) to 2.51 ng (11 times diluted). The acquisition was done with 100 repetitions that have been averaged afterwards leading to an effective acquisition time of 2.14~sec. 

To avoid misinterpretations of image artifacts as particle signal, each sample has been moved through the FOV within the experiment by a robot to 11 defined positions on the $x$ axis. Thus, a moving dot in the image can be identified as the sample. All time frames during the robot movement were neglected as the robot induces strong disturbances in both receive paths.

\subsubsection{Sample Volume Impact at Constant Iron Content}
To investigate the influence of dilution with a constant iron content, a 10~ng LS-008 sample is placed into the scanner and the previously described motion experiment is repeated. Then, the sample is diluted by adding deionized water in the sample chamber and the experiment has been repeated. Thus, the iron content was kept constant but was spread within the FOV. The measured volumes reached from 1~\textmu L to 128\,\textmu L increasing in stpdf of factor two.

\subsubsection{\textit{In Vivo} Experiments}

The \textit{in vivo} experiment using a healthy mouse were carried out using a similar workflow as outlined in\,\cite{Kaul2015} and\,\cite{werner2016geometry}. The experiment was approved by the local committee on animal protection (Beh\"{o}rde f\"{u}r Gesundheit und Verbraucherschutz, Freie und Hansestadt Hamburg, Nr. 42/14). All experiments were carried out in accordance with the relevant guidelines and regulations. The mouse was anesthetized using isoflurane and put on the mouse bed into a preclinical 7T MRI scanner (Bruker Clinscan). An isotropic gradient echo measurement sequence with 0.25~mm slice thickness and 128$^3$ voxels was measured with respiratory gating for reduced motion artifacts.

After the successful MRI measurement, the mouse was moved within the animal support unit to the MPI scanner. After initial background measurement, three different boli of LS-008 MPI-tailored contrast agent\cite{Ferguson2015}  were sequentially injected via a 27 G catheter into the tail vain. The three applied boli each had a total volume of 10~\textmu L and different total iron amounts of 0.514~\textmu g, 5.14~\textmu g, and 51.4~\textmu g. The time between the bolus injections was between 5 and 10 minutes.
MPI measurements were performed with the same imaging parameters as used for the \textit{in vitro} experiments. The FOV of $24\times 24 \times 12$\,mm$^3$ was placed covering the heart of the mouse. No averaging was done leading to a temporal resolution of 46 volumes/s (21.4~ms per frame).

\section{Results}
\begin{figure}
	\centering
	\includegraphics[width=\textwidth]{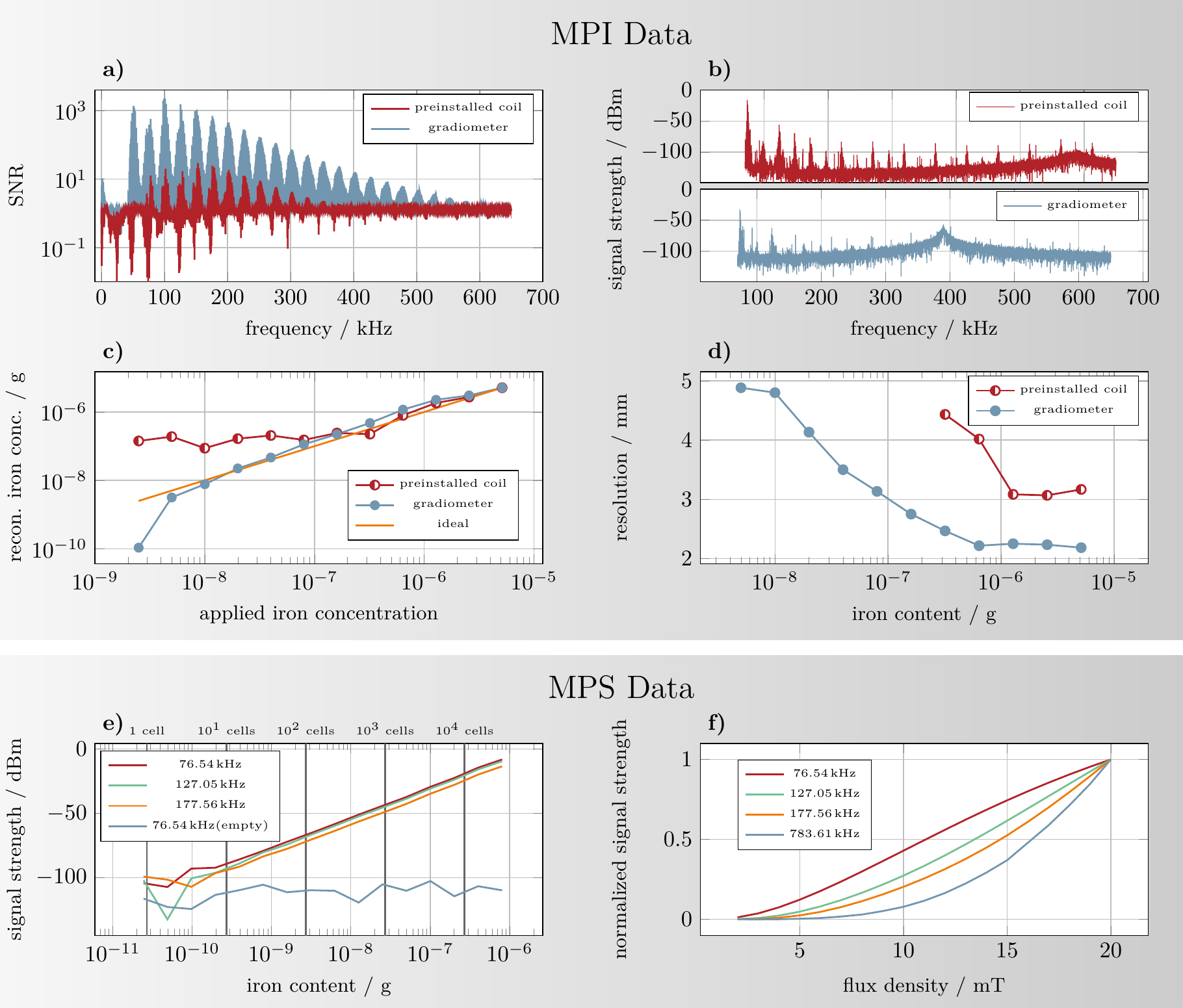}
	\caption{a) SNR of the system matrix used in the \text{in vitro} experiments for both receive coils as a function of frequency. b) Empty measurements of the installed receive coil and the gradiometer. The plotted MPI signal is given in dBm calculated in relation to a $50\,\Omega$ impedance. c) Calculated iron content determined by integration of the reconstructed particle distribution shown in Fig. \ref{Fig.ResultsSensitivity} plotted against the applied iron content. d) FWMH of the reconstructed dot in $z$ direction (vertical) as a function of the concentration for both receive coils. Shown are only those concentrations where a dot could be clearly detected. e) Sensitivity of the MPS device \cite{Graeser2017a}. The system shows a linear scaling of the harmonics down to a total iron content of 192~pg for LS-008 particles. For orientation, the number of cells for an iron load of 27~pg per cell as has been reported in \cite{zheng2015magnetic} is drawn. f) Scaling of the signal intensity over drive field strength for 10~\textmu L undiluted LS-008 particles. All harmonics show a nonlinear behavior for low drive field strengths converging to a linear scaling for higher field strengths.}
	\label{Fig:ResultData}
\end{figure}
\subsection{MPS Measurement Results}
Figure \ref{Fig:ResultData} e) shows the scaling of the 3rd, the 5th, and 7th harmonic component in a double logarithmic scale. Additionally, the signal at the 3rd harmonic of several empty measurements are shown as a reference for the system noise at identical measurement parameters. The signal strength shows a linear slope down to 192~pg. In further dilutions the signal is influenced by the systems noise.  

In figure \ref{Fig:ResultData} f) the amplitudes of four different frequencies are shown. The lower frequency components show a nonlinear increase for low drive field values converging to a linear slope for high field values. The knee point of the curve shifts towards higher values for higher frequencies. For the 30~th frequency component the linear slope has not been reached within the measured drive field range. As the sensitivity directly scales with the signal intensity of the harmonics, this has to be taken into account when comparing the sensitivity of one system with other systems.
\subsection{SNR Comparison}
Next, we compare the SNR of an undiluted delta sample measured with the preinstalled and the gradiometer coil. The results are shown in Fig.~\ref{Fig:ResultData} a). It can be seen that the SNR of the gradiometer is much stronger. But it can also be observed that the SNR difference is not constant but varies over frequency. Therefore, we calculated the maximum SNR around each harmonic and analyzed the ratio of both which can be interpreted as a gain factor. 

Due to the differences in the filter stages this gain factor varies between 120 within the blockband of the filter and 20 within the passband of the filter. Thus, we can expect the detection limit difference to be between one and two orders of magnitude.

\subsection{Empty Spectrum Comparison}

We first analyze the spectra of both the preinstalled and the developed gradiometer coil with an empty scanner bore. The results are shown in Fig. \ref{Fig:ResultData} b). It can be seen, that the installed coil shows strong signals around the pure harmonics (multiples of the excitation frequencies $f_x = 24.50$~kHz, $f_y = 26.04$~kHz, and $f_z = 25.25$~kHz. At each harmonic several side bands occur that are characteristic for a typical 3D MPI spectrum measured with SPIONs. This implies that the disturbing background signal can only be explained by magnetic material within the scanner that is excited by the 3D MPI Lissajous sequence.

Using the dedicated gradiometer coil the background signal is substantially reduced. While some background signal is still present around $75$~kHz and $125$~kHz, the background signal is much less pronounced around the other harmonics. The reason that some harmonics are still present can only be explained by a close distance of the distorting material to the receive coils such that the cancellation turns of the gradiometer see a different signal than the receiving turns. We note that the amplification of the magnetic moment is higher in the gradiometer receive chain which  can not be seen in this plot.

\subsection{Dilution Series}

\begin{figure}[!tb]
	\centering
\tabcolsep1pt
\includegraphics[width=\textwidth]{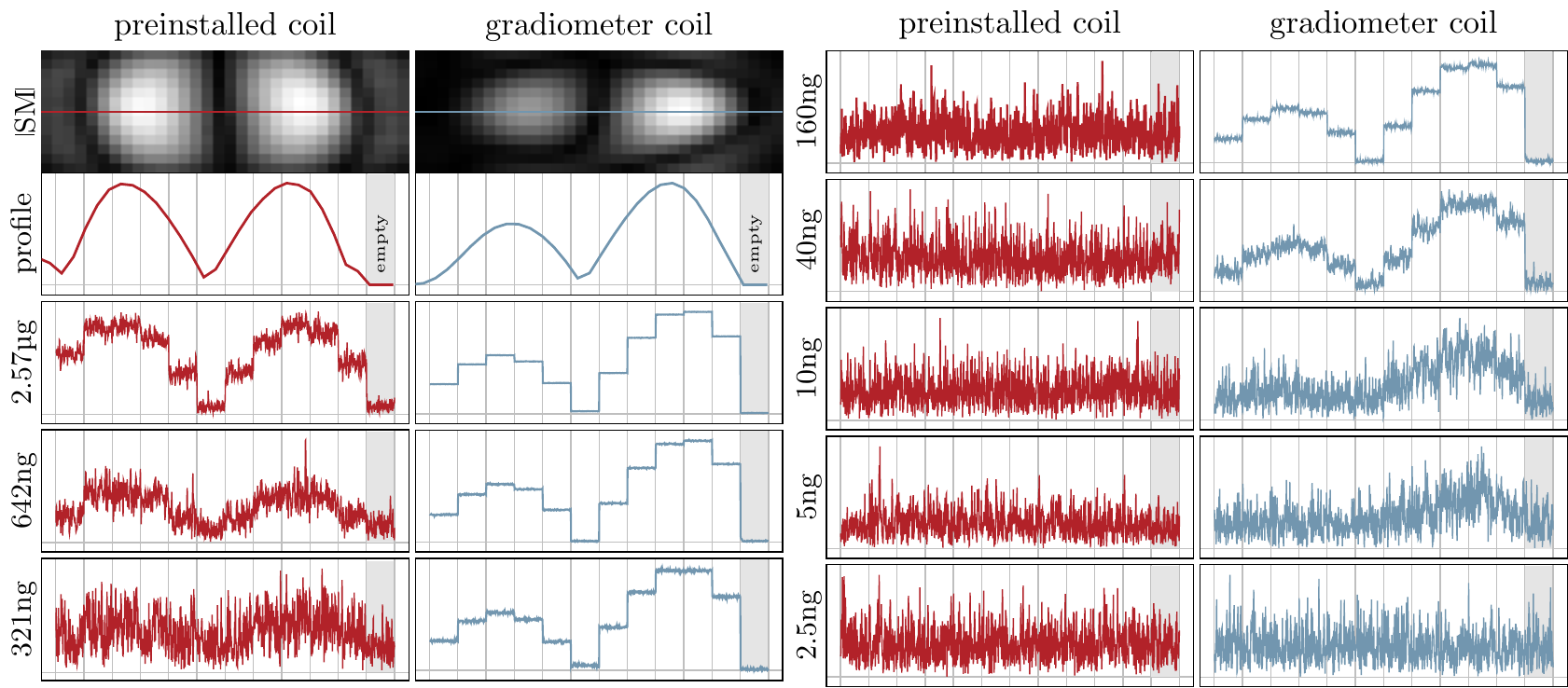}

\caption{Raw data analysis. The first image shows the absolute value of the SM component with the highest SNR for the preinstalled coil and the gradiometer coil. As the dilution samples are moved on the x axis, the resulting harmonic magnitude over time are expected to match the profile of these patterns shown in the second plot. The dilution raw data plots contain the unprocessed time data of the selected harmonics for the dilution samples at 11 positions in the FOV and a 12th position outside the scanner bore. Thus, the last part of the plot is actually an empty measurement. It is visible that the gradiometer is misplaced slightly causing a non-symmetric influence of the coil sensitivity.}
\label{Fig:ConcSeriesRawData}
\end{figure}

To determine the detection limit for each receive chain we used the dilution series and analyze the harmonic with the strongest signal. These occurred around 150~kHz for the preinstalled send/receive coil and around 100~kHz for the dedicated gradiometer coil. Fig. \ref{Fig:ConcSeriesRawData} shows the raw data signals of these harmonics for 100 subsequent frames at all 11 measured positions for selected dilutions in the FOV and additionally for an empty scanner. The asymmetric shape of the gradiometer pattern is caused by a slight misplacement of the coil within the FOV. As this misplacement is included in the system matrix, it has no effect on the reconstructed images. For the installed coil, one can observe that the detection limit is located between 160~ng -- 321~ng, since for the latter one can still distinguish the background signal from the signals at the first and the last spatial position. For the gradiometer coil the detection limit can be found to be between 2.5~ng -- 5~ng, since the signal at the third position is still observable in the raw data for the 5~ng sample.

\begin{figure}[tbh!]
\tabcolsep1pt
\centering

\includegraphics[width=\textwidth]{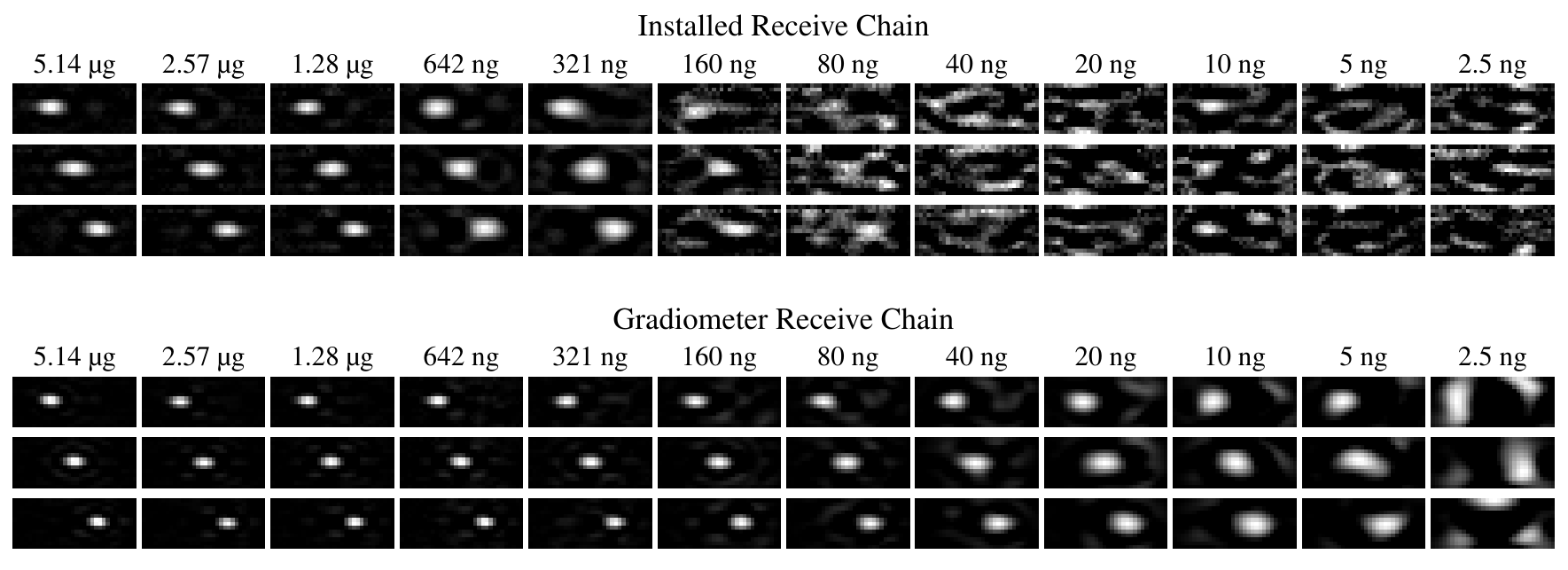}
\caption{Reconstructed image series of the moving particle sample for the preinstalled receive chain and the gradiometer receive chain. Each image is normalized to its current maximum intensity. With decreasing iron content the reconstructed sample appears blurred which is caused by fewer frequencies involved in the reconstruction due to the adapted SNR threshold. For the preinstalled receive chain the detection limit is reached at about 160~ng iron content. For the gradiometer receive chain the detection limit is improved by a factor of 32 to 5~ng iron content.}
 \label{Fig.ResultsSensitivity}
\end{figure}

We then reconstructed the raw data in order to analyze if the SNR of the measurement data was sufficient to actually calculate a spatially resolved image. The results are shown in Fig. \ref{Fig.ResultsSensitivity} for both applied receive coils. Three time points with the sample on the left, in the center, and on the right are shown. Each image was scaled for its own maximum intensity to visualize the wide contrast range. It can be seen that the dot is clearly  visible for high iron contents. With decreasing iron content the image quality degrades. One can first observe that the size of the dot gets larger which can be explained by the fact that the number of frequency components actually contributing to the reconstruction result decreases. At very low iron contents one can observe artifacts that can be explained by noise and remaining background signal that are reconstructed.

The detection limit for which it is still possible to reconstruct a spatially resolved image is at about 160~ng for the installed coil and about 5~ng for the gradiometer coil. This is in good agreement with the raw data analysis carried out before.

In order to investigate the linearity of the reconstructed particle concentration in dependence of the used iron content, the signal around the dot within the reconstructed particle distribution is integrated to obtain the total number of particles in the FOV. This is done since the particle concentration within a specific voxel is underestimated by the blurring effect leading to a spread of signal. The calculated iron volume is plotted against the applied iron content in Fig. \ref{Fig:ResultData} c). From this plot one can derive the detection limit of about 160~ng for the installed coil and 5~ng for the gradiometer coil. For lower iron amounts the signal level reaches a plateau which is not plausible. 

In order to illustrate the relation between the SNR of the measurements and the spatial resolution after reconstruction, the full-width at half maximum (FWHM) of the reconstructed dots in vertical direction is calculated in a subpixel accurate fashion and plotted against the applied iron content (see Fig. \ref{Fig:ResultData} d)). As one can see, the gradiometer receive coil provide in general a better spatial resolution. 
At lower iron content the resolution decreases in a monotonic fashion and drops from 2.1~mm at 5.14~\textmu g to 4.8~mm at 5~ng.

\subsection{Sample Volume Impact}

\begin{figure*}[tbh!]

\includegraphics[width=\textwidth]{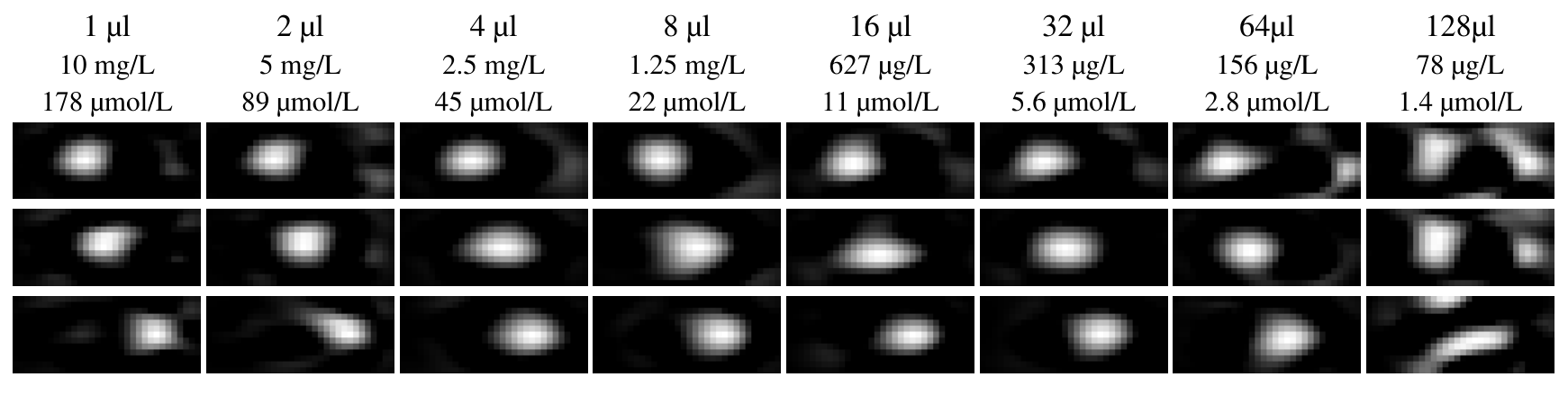}
\caption{Reconstructed image series of moving particle samples of different concentration. Each sample has the same iron mass of 10~ng but a volume ranging from 1~\textmu L until 128~\textmu L. The resulting concentrations range from 10~mg/L -- 78~\textmu g/L (178~\textmu mol/L -- 1.4~\textmu mol/L). The detection limit is reached at about 156~\textmu g/L (2.8~\textmu mol/L).}
 \label{Fig:VolumeSeries}
\end{figure*}

After determining the total amount of iron that can still be reconstructed with the gradiometer receive coil we next investigate the minimal iron concentration that can be detected. We note that this highly depends on the applied gradient strength of the selection field. In Fig. \ref{Fig:VolumeSeries} the reconstruction results of 10~ng iron mass LS-008 particles within volumes ranging from 1~\textmu L to 128~\textmu L at three different positions are shown. 
It can be seen that until about 64~\textmu L sample volume corresponding to an iron concentration of 156~\textmu g/L, the moving sample can still be detected, marking the lowest iron concentration published so far. The 64~\textmu L sample volume corresponds to an ellipsoid of $6.25 \times 6.25 \times 3.26$~mm$^3$ outer diameter. This is in a similar range as the size of the FFP convolution kernel for 2~T/m and nanoparticles of 25~nm core diameter \cite{Rahmer2009}.

\subsection{\textit{In Vivo} Experiments}

\begin{figure*}[tbh!]
\tabcolsep1pt
\centering
\includegraphics[width=\textwidth]{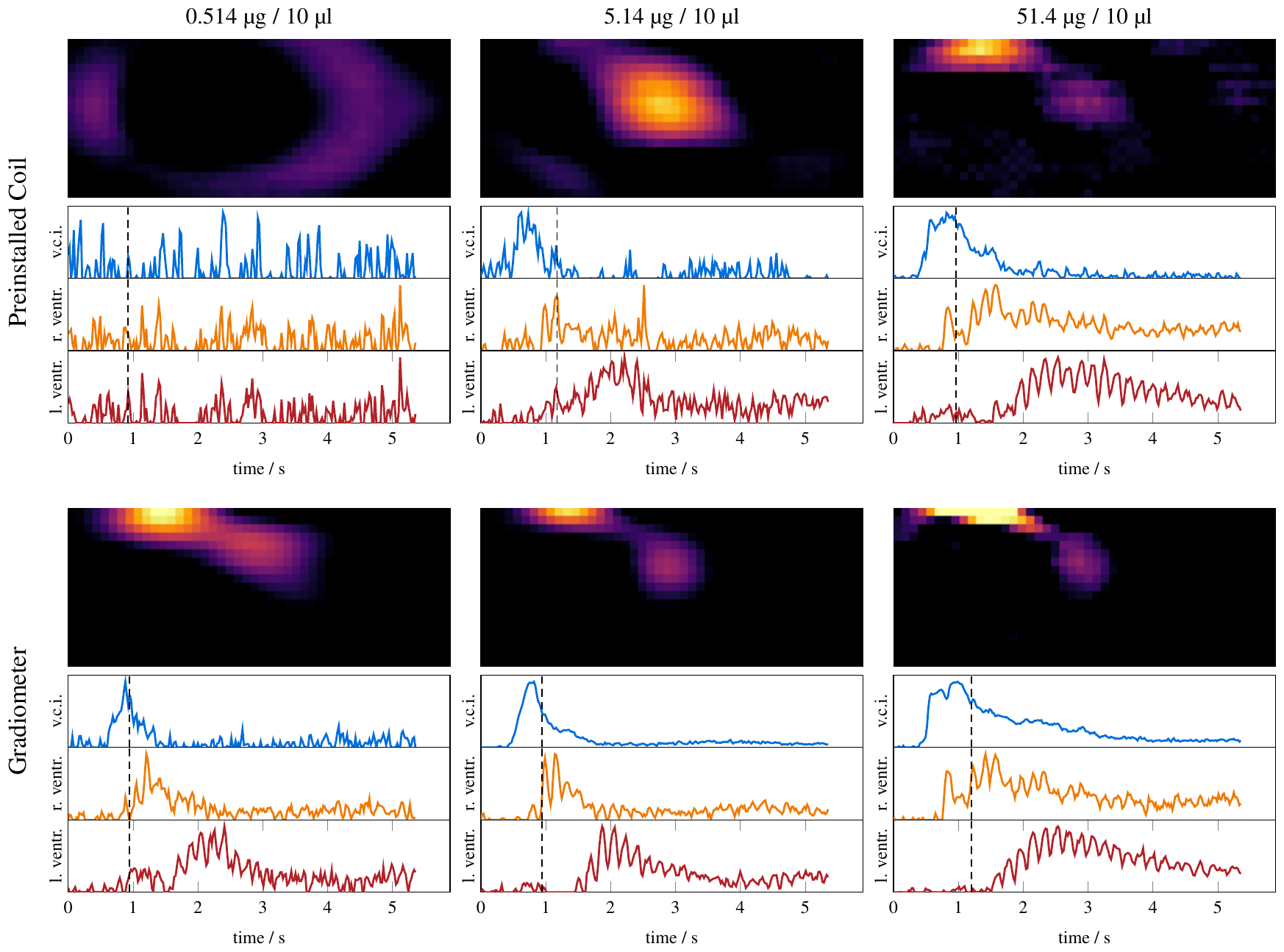}
\caption{
\textit{In vivo} reconstruction results of the mouse experiment with three different boli: 514~ng / 10~\textmu L (left), 5.14~\textmu g / 10~\textmu L (middle),
51.4~\textmu g / 10~\textmu L (right). The results for the preinstalled coil are shown at the top while the results obtained with the gradiometer coil are shown below. For each coil/bolus, a sagittal slice at the time where the SPIONs enter the mouse heart is shown. Additionally, temporal profiles at the vena cava inferior, the left ventricle and the right ventricle are presented.
}
 \label{Fig:InVivoComparison}
\end{figure*}

The results of the \textit{in vivo} experiments using different applied iron contents are presented in Fig.
\ref{Fig:InVivoComparison}. Shown are sagittal slices of 0.7\, mm slice thickness  at the time point where the bolus enters the heart via the right atrium. In addition, three time profiles in different voxels placed in the vena cava inferior, the right ventricle, and the left ventricle are shown in order of appearance.
At the lowest applied iron content of 0.514~\textmu g, the bolus can only be seen when using the gradiometer coil, marking the lowest dose \textit{in vivo} image so far. Higher dosages can be detected by both coils. When comparing the 51.4~\textmu g data of the installed coil with the 5.14~\textmu g data of the gradiometer, one can see that the latter is of better quality while the 514~ng data is of worse quality. Thus, as expected from the \textit{in vitro} experiments, the gain of the gradiometric receive coil is between 10 and 100.

\begin{figure*}[tb]

\includegraphics[width=\textwidth]{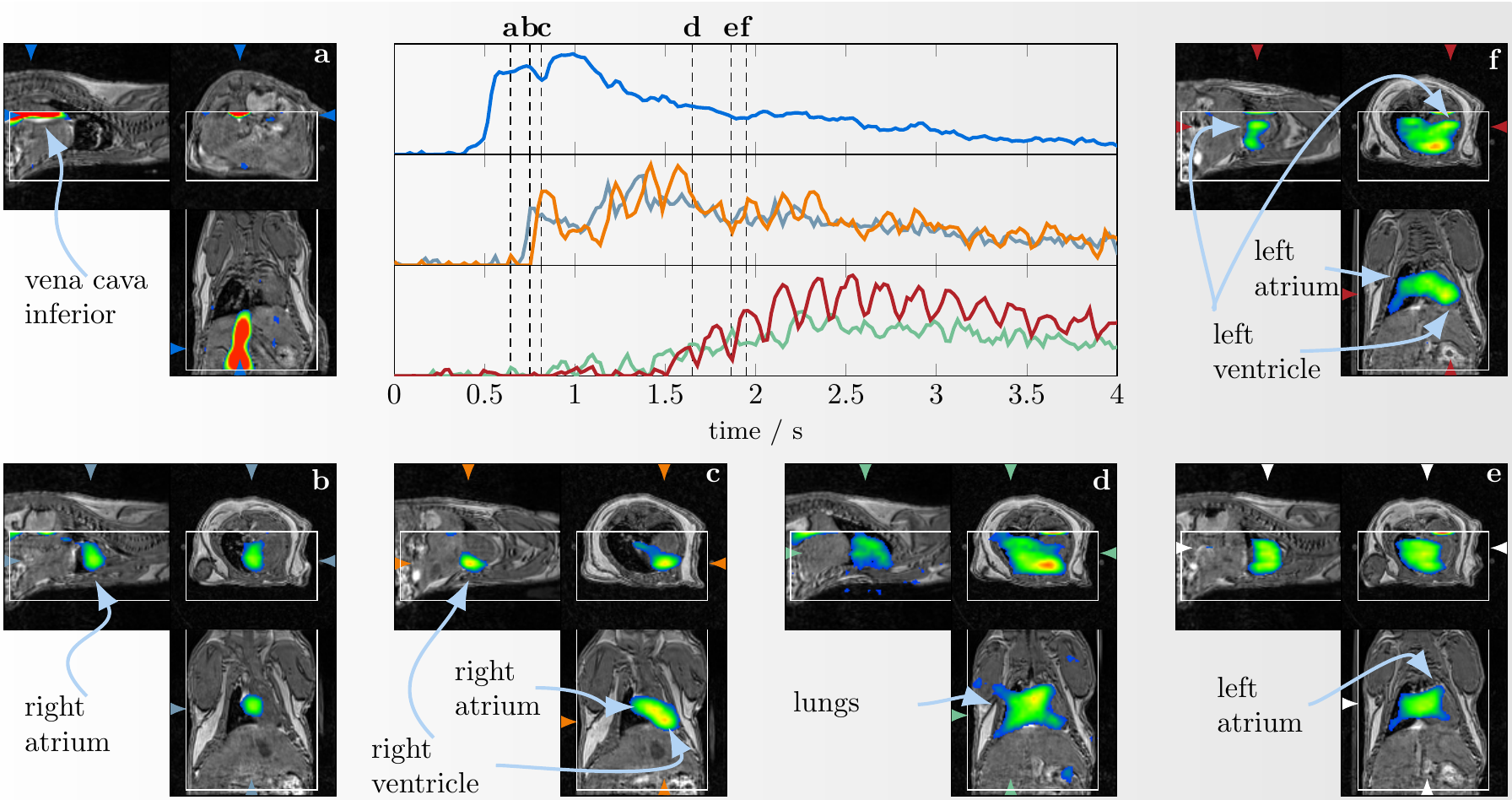}

\caption{4D \textit{in vivo} reconstruction results of the bolus experiment with the highest concentration (51.2~\textmu g / 10~\textmu L) at a time resolution of 21.41~ms. Shown are selected image fusions that reveal how the bolus passes through the mouse heart. The white box inside the images indicate the MPI FOV. Additionally, at the center the temporal progression of the signal in different selected structures (vena cava (top), right atrium and right ventricle (middle),  left atrium and left ventricle  (bottom)) are shown.}
\label{Fig:Fusion}
\end{figure*}

Next, we take a closer look at the reconstruction results obtained for the highest injection (51.4 \textmu g) measured with the gradiometer receive coil. In Fig. \ref{Fig:Fusion} selected slices of the reconstructed 4D MPI tomograms are shown. The images are shown ontop of the MR images providing an anatomical reference. In addition to the distribution images, Fig. \ref{Fig:Fusion} contains temporal profile plots of the signal within different image voxels placed at different structures within the mouse heart. The MPI data is plotted without averaging resulting in a temporal resolution of 21.41~ms.
As one can see, the bolus first appears in the vena cava inferior. It then enters the heart via the right atrium and 65~ms later, it fills the right ventricle. About 1~s later after passing the lungs and the 4 lung veins, the bolus arrives in the left atrium and finally, in the left ventricle. The temporal sequence of the bolus track can be directly seen in the profile plots. As to be expected, the signal in the atrium and the ventricle is out of phase since the cardiac oscillations have shifted maxima.

\section{Discussion}

We have shown in this paper, that MPI is a highly sensitive imaging method capable of detecting down to 5~ng iron content in only 2.14~s acquisition time \textit{in vitro} (100 averaged frames). When comparing the value to previous published images the ones most close are those published in \cite{zheng2015magnetic} where a limit of 27~ng has been reached in 20~s acquisition time. The authors used only two frequencies for reconstruction and extrapolated this value down to a 5.4~ng detection limit based on the assumption that the reconstructed image had an SNR of about 5. In our experience, such an extrapolation is not accurate when using a regularized reconstruction that removes noisy frequency components. In our exemplary data we were able to adjust the SNR of the 5~ng reconstruction to different values ranging from 1 to 100 and still were not able to detect the 2.5~ng sample.

Our dilution experiments of the 10~ng sample show that despite the lower concentration the dot is still visible until the volume of the sample is bigger than the sensitive area around the FFP. The lowest iron concentration was 156~\textmu g/L (2.8~\textmu mol/L) which is currently the lowest concentration reconstructed as an image. 

Our \textit{in vivo} experiments reproduce the finding of\,\cite{Weizenecker2009} proving that MPI is capable of imaging the flow of an intravenously injected bolus through the heart of a mouse and differentiate the larger anatomical structures like the vena cava inferior, the right and the left atrium and the right and the left ventricle. The spatial resolution of our images is comparable to the images published in\,\cite{Weizenecker2009} despite of a lower applied gradient strength of 2~T/m instead of 5.5~T/m. The boli applied by Weizenecker et al. had iron masses of 14~\textmu g and 56~\textmu g which are comparable to the 5.14~\textmu g and 51.4~\textmu g applied in our experiments \cite{Weizenecker2009}. In addition, we applied and detected a 514~ng bolus marking an improvement of about 27 times compared to the lowest previous published value of 14~\textmu g\cite{Weizenecker2009}.

We have proposed guidelines for determining the detection limit of MPI scanners that are based on a dilution series circumventing the need for extrapolations. One key element of the protocol is to move the object to different positions in the FOV to avoid identifying artifacts spuriously as particles. In order to make the procedure independent of the encoding scheme, the particles should be located at a very small volume. 

Although, the developed coil in combination with the noise matching network is already highly optimized, there is still room for improvement in sensitivity. In particular, the LNA used in our work is not fully optimized yet. Weizenecker at al. reported on
a liquid cooled J-FET-based LNA having an input noise voltage of $80\,\text{pV}/\sqrt{\text{Hz}}$ which is potentially a factor of 5.5 better than our LNA.  However, as the input capacitance is reported to be 1\,nF (3 times higher than the LNA used in this work), the improvement for the effective sensitivity will be about $\frac{5.5}{\sqrt{3}}\approx3.2$. Since we used only frequencies in the range 47.4~kHz-227.5~kHz for the reconstruction of the 5~ng sample, it may additionally be possible to limit the analog bandwidth of the receiver to a small frequency range allowing for further noise reductions.

The developed gradiometer coil was tailored to fit the support unit of a mouse bed and had an inner diameter of 40~mm. When scaling up the diameter to 20~cm, which would be a realistic value for human surface coils, one would loose about a factor of 11.2 in sensitivity by the geometric differences. This would lead to a detection limit of about 50~ng iron content and about 1.56~mg/L (28~\textmu mol/L) iron concentration at a gradient strength of 2~T/m when using the same scanning parameters as in this work. This would already be sufficient for vascular imaging as required for interventional applications. When exploiting the aforementioned technical opportunities and when using an FFL encoding scheme, the sensitivity could be improved by a factor of 10-100 which would enrich the range of possible medical applications.

\section{Appendix}
\label{Appendix}
In this appendix, we derive the approximitive  sensitivity law outlined in Section \ref{Sec:SensitiviyLaws}. We consider the coil sensitivity of a solenoid at the center ($\vec{O}$) to be
\begin{align}
p_x(\vec{O})=\frac{N\mu_0}{l\sqrt{\frac{4R^2}{l^2}+1}}.
\end{align} 
The inductance of a short solenoid can be calculated to be\cite{Wheeler1928}
\begin{align}
L=\frac{\mu_0 N^2 \pi R^2}{l+0.9 R}.
\end{align}
The coil efficiency $\rho$ can then be written as 
\begin{equation}
\eff=\frac{p_x(\vec{O})}{\sqrt{L}}=\frac{\frac{N\mu_0}{l\sqrt{\frac{4R^2}{l^2}+1}}}{\sqrt{\frac{N^2\mu_0 A}{l+0.9 R}}}.
\end{equation}
With $l=1.3R$ this can be further simplified to 
\begin{align}
	\eff=\frac{\frac{N \mu_0}{2.385 R}}{N\sqrt{\frac{\mu_0\pi R^2}{2.2R}}}=\frac{\sqrt{2.2\cdot\mu_0}}{2.385 R \sqrt{R\pi}}\propto \frac{1}{\sqrt{R^3}}.
\end{align}
Thus, the sensitivity scales inversely with $R^{1.5}$.

\section{Acknowledgements}
The authors would like to thank Reinhard Schulz and Dirk Steinhagen for their help in design and construction of the coil mounting. 

The authors would like to thank Peter Ludewig, Nadine Gdaniec and Tobias Mummert for their assistance during the measurements. 

M.G., A.v.G., T.F. and T.M.B. thankfully acknowledge the financial support of the German Research Foundation (DFG) and the Federal Ministry of Education and Research (BMBF) (Grant Numbers BU 1436/10-1 and 13GW0069A).

T.K., P.S. thankfully acknowledge the financial support by the German Research Foundation (DFG, grant numbers KN 1108/2-1 and AD 125/5-1).

Work at UW was supported by NIH grant R42 EB013520-02A1. K.M.K also acknowledges the Alexander
von Humboldt Foundation for the 2016 Forschungspreis.

\bibliography{ref} 
\section{Author Contributions}
M.G., P.S., T.F., T.M.B., H.I., G.A. and T.K. contributed to critical discussions toward experimental parameters. M.G. designed and constructed the gradiometric receive coil as well as the receive chain. T.F., P.S., M.G. T.K. performed the MPI sequences. P.S. and T.K. performed the image reconstructions. M.K. performed the MRI sequences as well as the animal handling. A.v.G. and M.G. performed the MPS measurements.  K.M.K. manufactured and provided experimental optimized tracer material for both the in vitro and in vivo measurements. M.G., P.S. and T.K. contributed to writing the paper. All authors reviewed the manuscript.

\section{Additional Information}
\subsection{Competing financial interests:}
The authors state no competing financial interests.
\end{document}